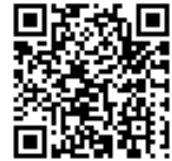

# Framework of Social Customer Relationship Management in E-Health Services

**Muhammad Anshari and Mohammad Nabil Almunawar**

Universiti Brunei Darussalam, Brunei Darussalam

___________________________________________________________________________________

**Abstract**

Healthcare organization is implementing Customer Relationship Management (CRM) as a strategy for managing interactions with patients involving technology to organize, automate, and coordinate business processes. Web-based CRM provides healthcare organization with the ability to broaden service beyond its usual practices in achieving a complex patient care goal, and this paper discusses and demonstrates how a new approach in CRM based on Web 2.0 or Social CRM helps healthcare organizations to improve their customer support, and at the same time avoiding possible conflicts, and promoting better healthcare to patients. A conceptual framework of the new approach will be proposed and highlighted. The framework includes some important features of Social CRM such as customer's empowerment, social interactivity between healthcare organization–patients, and patients-patients. The framework offers new perspective in building relationships between healthcare organizations and customers and among customers in e-health scenario. It is developed based on the latest development of CRM literatures and case studies analysis. In addition, customer service paradigm in social network's era, the important of online health education, and empowerment in healthcare organization will be taken into consideration.

___________________________________________________________________________________

## Introduction

Nowadays, the use of Information and Communication Technologies (ICT) has widely spread in many aspects of our lives and its impact is growing in various sectors, including the healthcare sector. The utilization of ICT, especially the Internet, in the healthcare sector is frequently referred as electronic health or e-health is basically to improve healthcare management for mutual benefit between patients and healthcare providers. One important aspect of e-health is how to manage relationships between a healthcare provider and its customers (patients) in order to create greater mutual understanding, trust, and patient involvement in decision making.

A good relationship between a healthcare provider and its customers does not only improve customer's satisfaction, but also helps in fostering effective communications between them, which may help to improve their health and health-related quality life and this more effective in chronic disease management (Arora, 2003). In addition, the role of ICT in managing relationship with customers in ways that can retain them is very important. ICT helps in fostering customer self- service and empowers them which allow an organization to reduce costs by handling an increasing number of consumer's transactions effectively. In addition, meeting the customers' expectations that becoming increasingly mobile poses significant challenge on how effectively an organization attach with their customers. Thus, ICT strategies should be aligned with business strategy focusing on customer satisfaction in acquiring, managing, and maintaining them for long time engagement known as Customer Relationship Management (CRM), which can be viewed as strategy to attract new customers coming to an organization, retaining them throughout the entire lifetime of a relationship, and extending other services or products to the existing customers.

However, the gap between managing CRM and customer care needs make fulfilling the needs through more complexes. The complexity increases due to the changing of customer behavior which driven by technology advancement. We are witnessing the acceptance of a second generation of web based communities such as wikis, blogs, and social networking sites which aim to facilitate creativity, collaboration, sharing among users rather than just for sending/receiving emails and retrieve some information. It is important to note that with Web 2.0 users can own data and exercise control over their data (Hinchcliffe, 2006). When a customer becomes more empowered through owing the data on the Web 2.0., their relationship with an organization that serves them will be enhanced.

The main goal of this paper is to introduce a promising future research direction which will shape the future of health informatics. In this paper we will discuss and demonstrate how a new approach in CRM based on Web 2.0 known as Social CRM or CRM 2.0 will help healthcare providers improve their customer services, avoiding conflict, and promoting health education to patient which can lead to improve their health literacy and customer satisfaction. A conceptual framework of such new approach will be proposed and highlighted, and supported with a compelling example. The structure of the paper is as follows. In sections 2, we discuss about CRM and Social CRM. In section 3, we propose our conceptual model for CRM within e-health system. In section 4 is the conclusion.

**Literature Review**

The role of ICT in fostering effective management in a healthcare organization is inline with the definition of e-health. Eysenbach (2001) defines e-health as an emerging field in the intersection of medical informatics, public health and business, referring to health services and information delivered or enhanced through the Internet and related technologies. In the context of Europe, the concept of e-health is used to describe the application of ICT across the whole range of functions which, one way or another, affect the health of citizens and patients. The term e-Health covers a range of technological areas. Furthermore, World Health Organization (WHO) defines e-health as a cost effective and secure use of ICT in support of health and health related field, including healthcare service, health surveillance, health literature and health education, knowledge and research (WHO, 2005).

As a matter of fact, an e-health is not going to replace whole existing healthcare services. It is an extendable system which its features and services can improve quality of service, provide effectiveness in process, and convenience for patients who rely heavily with the ICT. E-health is expected to provide more comprehensive and reliable healthcare in serving patients. For instance, it offers more options for patients to consult either physically meet a doctor or virtually meet a doctor online. The study proposed a general framework how Social CRM embedded into e-health systems can improve customer satisfaction and health literacy of individual. We proposed the features of Social CRM as social networks, empowerment, and online health educator (see figure 1). These features are expected to achieve health literacy and at the same time improving customer satisfaction as defined at e-health's definition. The features are proposed based on the literature analysis and recent case study of managing customer in healthcare organization.

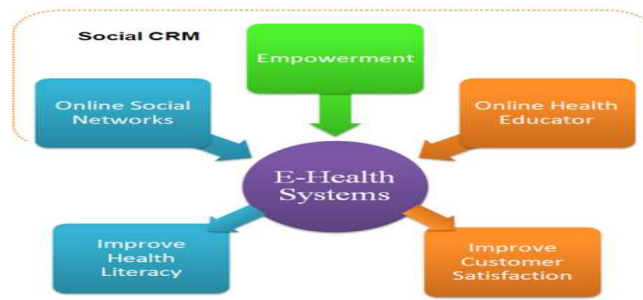

**Fig 1. E-Health with Features of Social CRM**

Social CRM, which is based on Web 2.0 should accommodate feature of social networks. Social networks in a healthcare organization can be incorporated in a strategy to win customer hearths and minds through better understanding what customers (patients) want from their services. Through conversation in social networks, customers feel free to give appraisal or criticize the organization. Social networks can also be media for sharing the experiences of the patients with the same condition.

Empowerment emerges from the change of philosophy that nowadays an individual have better control and responsibilities in taking appropriate decision of their own health. Empowerment can be used in building customers' trust. Finally, online health education is critical component in any e-health initiative. The availability of person in charge who support the online customers play critical role in order to achieve the goal of e-health. These three features will be discussed in the following section to portray the outcome to improve health literacy and customer satisfaction.

**Customer Relationship Management**

CRM is a broad term and widely-implemented strategy for managing interactions with customers. CRM use ICT to organize, automate, and synchronize business processes—principally customer service, marketing, and sales activities. The overall goals are to find, attract, and win new customers, nurture and retain those the company already has, entice former customers back into the fold, and reduce the costs of marketing and customer service (Galimi, 2001).

In practice, many see CRM is merely a technology for improving customer service which may lead to a failure when implementing it. CRM initiatives must be seen as a strategy for significant improvement in services by solidifying satisfaction, loyalty and advocacy through information and communication technology. As such, matters pertaining related to people such as customer behavior, culture transformation, personal agendas, and new interactions between individuals and group must be incorporated in CRM initiatives. Therefore, organization needs to understand that behavior and expectation of customers (patients in healthcare organizations) which continue to change overtime. Consequently, CRM must address the dynamic nature of patients' needs and hence adjustments strategies embedded in CRM are required.

Greenberg (2004) defined CRM as a philosophy and a business strategy supported by a system and a technology designed to improve human interactions in a business environment. It is an operational, transactional approach to customer management focusing around the customer facing departments, sales, marketing and customer service. Early CRM initiatives was the process of modification, culture change, technology and automation through use of data to support the management of customers so it can meet a business value of corporate objectives such as increases in revenue, higher margins, increase in selling time, campaign effectiveness, and reduction in call queuing time. While nowadays, CRM is designed to engage customers in a collaborative conversation in order to provide mutually beneficial value in a trusted and transparent business environment. When the strategy is created and developed into a proper planning and a good selection of supporting technology, the CRM is expected to have the ability to manage those relationships.

Although the development of CRM has been mature, there are many challenges in adopting CRM for healthcare organizations. Due to the complexity of the business nature in healthcare there are many issues dealing with patients that must be considered. A healthcare is undergoing a paradigm shift, from 'Industrial Age Medicine to Information Age Healthcare' (Smith, 1997). This 'paradigm shift' is shaping health systems (Haux et.al, 2002). It is also transforming the healthcare-patient relationship (Ball, 2001). For example, World Wide Web has changed the way the public engage with health information (Powell et al., 2003). According to Pew Internet and American Life Project, large shares of Internet users say that they will first use Internet when they need Information about healthcare (Pew Internet, 2005). People begin to use Internet resources for research on the health information and services that they are interested in using it. ICT creates an environment where patients can explore clinical records and health education programs anytime anywhere.

CRM strategy must be aligned to the organization's mission and objectives in order to bring about a sustained performance of business objectives and effective customer relationships. The organization must adopt customer's perspective and work on developing a comprehensive planning write up and specific business objectives. The strategies should be laid down in such a way so that they provide benefits to the company and customers, shorter cycle times, greater customer involvement in service development and reduce operation costs by redesigning business process that eliminates work which does not add value to customers (Thompson, 2006).

Relationship between a healthcare organization and its customers are varies with different in the level of complexities. In order to understand the features of Social CRM in managing relationship patient-organization, we offer the following case between a patient whose name is Prita Mulyasari (Prita) with the Omni International Hospital Alam Sutera Tangerang, Indonesia (Omni). The case generated massive public attention channels through various media, including social network sites. We will use this case to motivate us in developing a new Customer Relationship Management (CRM) model to address relationship management between patients and healthcare providers that incorporate the latest development in Information and Communication Technology (ICT).

*Case # 1; Prita Mulyasari is a Tangerang Jakarta housewife and mother of two who was a patient at Omni International Hospital for an illness that was eventually misdiagnosed as mumps. Her complaints and dissatisfaction about her treatment which started as a private email to her friends in September 2008 were made public rapidly distributed across forums via online mailing lists. Once the email became public knowledge, Omni responded by filing a criminal complaint and a civil lawsuit against Prita. Then, verdict against Prita, at Banten District Court on May 13, 2009, she was sentenced to six years jail and fined 204 million rupiah (US$ 20,500). Support from a group on Facebook has attracted considered support as well as various Indonesian Blog. A mailing list and Facebook group called "KOIN UNTUK PRITA" (Coins for Prita) started raising money from people throughout Indonesia. People began collecting coins to*

*help Prita to pay the fine. Seeing the huge support for Prita, Omni International Hospital dropped the civil lawsuit. Significant pressure eventually led to Prita being released from detention on June 3, 2009 (Detik, 2008).*

While in the healthcare environment, healthcare organizations are challenged to acquire potential customers for the healthcare services, retaining them to use the services, and extending various services in the future. The case above shows how inappropriate respond could affect an organization in acquiring potential customers in the future. Therefore, to take the challenges, healthcare organization must consider to establish close of relationship with their patients offer convenience of services, and provide transparency in services through information sharing.

Therefore, the healthcare organization should perform re-engineering process to adapt their CRM strategy and tool in order to acquire potential customer coming for the service, and existing CRM which has been introduced within healthcare organization needs to be adjusted in order to accommodate the changing preference of customer due to technology advancement. (Anshari and Almunawar, 2011).

Figure 2 shows the business strategy on CRM. The model is a hybrid and having three key phases and three contextual factors; three key phases are customer acquisition, retention, and extension. And the other three contextual factors are marketing orientation, value creation, and innovative IT (Marketingteacher, 2010).

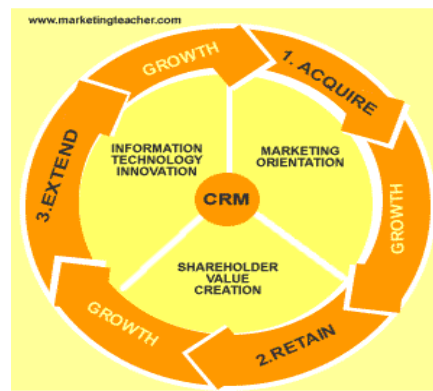

**Fig 2. Business Strategy and CRM Model (Marketingteacher.com)**

The customer acquisition is the process of attracting customers for the first of their purchase or use services. The customer retention is the customer return to us and uses the service for the second time. We keep them as our customer. The customer extension is introducing new product or service line to our loyal customers that may not relate to the original services or products. Growth the numbers of new or retain customers use the product or service through marketing orientation, value creation and innovative IT. When the hospital acquired a new patient for service through marketing orientation, the patient will determine the value of each activity received from the healthcare provider. However, it has not incorporated social network within the process of cycle times. This may affect to the process of retention and extension.

Refer back to the Prita's case above; from the perspective of CRM, the organization faced a serious problem in handling the customer's complaint. Inappropriate respond could cause the organization will loss customers in the future. Therefore, it is important to implement the right strategy in accommodating the customer's demand.

**Social CRM**

The rise of social media powered by Web 2.0 is changing how people send/receive, because everyone can create the message and share their opinions or experiences with each other. It is a revolution on how people communicate. It facilitates peer-to-peer collaboration and easy access to real time communication. Because much of the communication transition is organized around web based technologies, it is called Web 2.0 (Greenberg, 2009). Patients participate in these social network can share information about their diagnoses, medications, healthcare experiences, and other information. It is often in form of unstructured communication which can provide new insights for people involved in the management of health status and chronic care conditions.

Greenberg (2009) defined Social CRM as a philosophy and a business strategy, supported by a technology platform, business rules, processes, and social characteristics, designed to engage the customer in a collaborative conversation in order to provide mutually beneficial value in a trusted and transparent business environment. It's the company's response to the customer's ownership of the conversation. The term of Social CRM and CRM 2.0 is used interchangeably. Both share new special capabilities of social media and social networks that provide powerful new approaches to surpass traditional CRM.

Fabio Cipriani (2008) described the fundamental changes that Social CRM is introducing to the current, traditional CRM in term of landscape. It is a revolution in how people communicate, customers establish conversation not only with the service provider but it is also with others. Table 1 summarizes the difference of CRM 2.0 from CRM 1.0 based on type of relationship, connection, and how value generated. Relationship type in CRM 1.0 focuses on the individual relationship; Customer to Customer or Customer to Business whereas in CRM 2.0 offers the collaborative relationship and engage a more complex relationship network. Connection type in CRM 1.0 is limited view of the customer which affect to less informed customer, on the other hand, CRM 2.0 enable for multiple connections allow better understanding and more knowledgeable customer. CRM 1.0 of value creation is constricted from targeted messages, and CRM 2.0 offers diverse value creation even from informal conversation of customers within social networks.

Table 1: Comparison CRM 1.0 and CRM 2.0

| Type | CRM 1.0 | CRM 2.0 |
|---|---|---|
| **Relationship** | Focus on individual relationship (C2C, C2B) | Focus on collaborative relationship (engaging a more complex relationship network) |
| **Connection** | Limited view of the customer & his community preferences, habits, etc | Multiple connections allow better understanding of the customer and his community |
| **Generated Value** | Targeted messages generate value | Conversation generates value |

1. **Online Social Networks**

Online social networks are social relations or networks focuses on connection which enable them to build online profiling to initiate and connect with friends, friends of friends, or others who might share the same interest, ideas, activities, events, etc. Users can generate content online such as video, photo, music, online conversation, and other digital media. Popular types of social networking services are stating from email service, mailing list community, and instant messenger. Recently, social networking tools which is based on Web 2.0 technology such as Facebook, Twitter, MySpace, Friendster, LinkedIn, etc, have rapidly facilitating peer-to-peer collaboration, ease of participation, and ease of networking social.

Denis Pombriant from Beagle Research Group (2009) analyzed that customers has more control of their relationships with organization a lot more than they were in few years ago. It is because customers have accessed to new levels of education, wealth and information. The change is a social change that impacts all organizations. Customers share experiences about the service received from the organization through social networks. When they have good experiences with the organization, they can be informal representative that promoting the organization to their networking friends, the same case if they have unhappy experiences during their engagement with the organization, they can express their dissatisfaction to their friends on the social network which can generate distrust towards the organization. Public has built image from the effect of conversation taken place on social network that the hospital unprofessional in conducting medical activities which led to distrust to the service.

The Prita's case above proves that the online social networks nowadays play critical role for the success of CRM implementation. In CRM 1.0, customers heavily depended on the information from the organization simply because that time there was no mean to enable them making conversation between customers. However, customers now are making conversation and discuss even criticize the organization within their groups as what Prita did with her friend when complaining her dissatisfaction from the service of Omni. Therefore, it is necessary for the healthcare organization to understand and make sure they are listening what customer's expectation from their services.

2. **Empowerment**

The effort to attract a customer can take a month or even more but to lose one is easy. So there must be value of pleasing a customer, satisfying the customer's need, and build long lasting relationship between organization and its customers. Customer empowerment plays important role in building good relationship between an organization and its customers. Customer empowerment is increasingly important to help creating a competitive advantage and to compete and win in that market scenario. In general, empowerment is a characteristic of groups and individuals that energize them with the knowledge and confidence to act in their own behalf in a manner that best meets identified goals. In this sense, empowerment characterizes the manner in which patients and clinicians approach care, with mutual expectations, rights, and responsibilities. Empowerment represents a change in philosophy for both care providers and patients alike, requiring the former to abandon the authoritative control once held and the latter to assume a greater level of deliberate self involvement in the care process (Brennan, and Safran. 2005). On an individual level, empowerment is a social process of recognizing, promoting, and enhancing peoples' abilities to meet their own needs, solve their own problems and mobilize the necessary resources in order to feel in control of their lives (Gibson, 1991). For consumers to be fair and equal participants in empowered partnerships with clinicians requires that they have adequate knowledge; set realistic goals; access systematic.

In this regards, healthcare organizations should consider empowerment as part of their business strategy to win customers heart and minds by allowing them to act on their behalf. Customer must be given them the authority and responsibility to follow through in benefiting both parties. Empowerment allows the organization to delegate partially or full authority to customers, and trust them to act on their own benefit and responsibility. Empowerment gives them the necessary tools to resolve most questions faced by and reduce the number of their dissatisfaction. It also means that responsibility goes one step further to the customers.

Referred back to the Prita's case, when the patient felt that he was not able to access her medical record to seek second opinion, the patient is powerless in relation with the healthcare organization. Unfortunately, the feeling of powerless may create customers dissatisfaction and share dissatisfaction to her circles.

Likewise, in a healthcare organization, the authority to act is the set of resources that the patient has access to and the decisions they are permitted to make. At this point, e-health system can empower patients to have the ability in controlling his/her data. It will enable them to have personalized e-health service with Social CRM as the frontline of the system. The authorization and self-managed account/service are granted to access full or some parts of applications and data offered by the systems depending on the policy. The empowerment will benefit both healthcare organization as service provider and patients as users in the long run since the information and contents continue to grow. Furthermore, since all the information (medical records) can be accessed online everywhere and anytime, it will enable collaborative treatments.

3. **Online Health Educator**

Conventional wisdom holds that to increase loyalty, companies must delight customers by exceeding service expectations. Though Dixon et al. (2010) argued that a large scale study of contact centre and self service interactions, finds that to really win customer's loyalty is important to solve their problems. And nowadays, customers are enthusiastic with their dependency towards recent development of Web 2.0 namely Social Networking sites. In this regards, it is beneficial for any organization to consider the technology included in developing customer service's strategy. As a business, healthcare provider places in need of the same standards of customer service as other business organizations. The fact that customer service expectations in healthcare organization are high poses a serious challenge for healthcare providers as they have to make exceptional impression on every customer. In the competitive commercial healthcare environment, negative experiences, and poor service leads customers to switch healthcare providers because poor service indicates inefficiency, higher cost and lower quality of care.

These days, patients have more options in where they seek for care and how they interact with their healthcare providers. A great customer service can lead to major improvements in the healthcare system. Customer service is an essential requirement for providing high quality healthcare and for staying in highly competitive business (Stanton, 2009). Patients make choices about where they receive care based on service experiences and it is crucial for organizations to create an institutional ability to sense and respond, sympathy, and empathetically (Katzenbach, 2008). The case below explains how important the role of online health educator when the healthcare organization decides to implement e-health. The existence of online health educator will determine the level of satisfaction and at the same time educate the patient through online.

*Case # 2; I recently encountered something upsetting with the hospital. A few weeks ago, I phoned the hospital and asked to speak with a doctor at one of the specialist divisions. When the receptionist picked up the phone, I told her I wanted to speak with a doctor. To my surprise, she said I could not even leave a message as their office was about to close. (That was at 3.30 pm). When I requested again that I wanted to leave a message, she*

*still declined to accept it, repeating that they were about to close. I was puzzled at the attitude of the staff. Was she uncooperative because she was lazy or was it because she just could not care enough to accommodate my request? Hospital employees should be more caring and not disregard people who call in. What if it had been an emergency? This could really ruin the reputation of the hospital. (Brueitimes, 2009).*

The customer in the case above obviously demanded that there should be person in charge who responsible to respond to the service which facilitate offered by healthcare organization. When the organization promises to provide e-health service, it is no alternative for the organization unless they have to make sure that online health educator available as backbone of the Social CRM. It must be part of the strategy for the success of e-health implementation.

Healthcare organization's strategies should transform customer strategies and systems to customer engagement. The one is more focused on the conversation that is going on between organization and customer and the collaborative models that cutting edge companies are carrying out for customer engagement. Proactive strategies will improve customer services. And great customer support will increase loyalty, revenue, brand recognition, and business opportunity.

Reacting to problems after they happen is usually more expensive than addressing them proactively. Though the hospital have implemented all CRM technology and communication channels in the front end, they are not guarantee to be able to manage the relationship with their patients without involving right strategies such as manage conversation with the patients, and improve customer services.

At the fundamental of any organization is the organization knowledge of customers. When the organization manages resources well, it means they know customers well. They are able to customized services based on knowledge of their customers. People who are weak in direct communication can become more powerful in virtual communication such as email, instant messaging, video conferencing, etc. People who are introverts are seemingly able to communicate better, because they can plan or be better prepared before interacting with other party via ICT.

In the same way, e-health services which empower patients to have better knowledge and control over their health data, yet they need to communicate with healthcare organization. In order to achieve the goal, the presence of online health educator determines success or failure of the implementation. It should be part of the implementation strategy to ensure that there are groups of staff to ensure that e-health service is managed in professional ways. Any health activities which they can accomplish through online service, they may not come physically to the hospital. Though they do not come to the healthcare centre, yet they may need an assistant or guidance from online health educator when they face difficulties interacting with systems which they need to discuss with online health educator for clarification and interpretation such as medical data, online consultation, or asking for any online service. Presence of online health educator is the vital point in e-health instead of ICT as tool. They are expected to have skills to interpret medical data, able to guide patient go through technical systems, and know how to respond online queries properly. In short, online health assistant is available through online health informatics officer (like any other customer service in business/organization) who stands by assisting patient/family in make use of systems (Anshari et al., 2012)

**Discussion**

In response to the expectation of customers, we propose Social CRM features as front end to tie up the interface of e-health. It is significant to improve customer satisfaction and health literacy in healthcare service within e-health systems. Achieving health literacy is one of the main objectives of e-health as defined by WHO. Moreover, customer satisfaction is a key ingredient to engage customers in the long lasting relationship with the organization. Customer satisfaction will ensure that the healthcare organization has trusted and reliable service in healthcare processes. Therefore, based on the cases, we will elaborate the component of Social CRM undertake mutual benefits for both organization and customers.

The model operates in the area of healthcare organization–patient relationships inclusive with social networks interaction, and how they possibly shared information to achieve health outcomes. Figure 3 is a proposed model of Social CRM within e-health environment. It offers a starting point for identifying possible theoretical mechanisms that might account for ways in which Social CRM provides one-stop service for building relationship between healthcare organization, patients, and community at large.

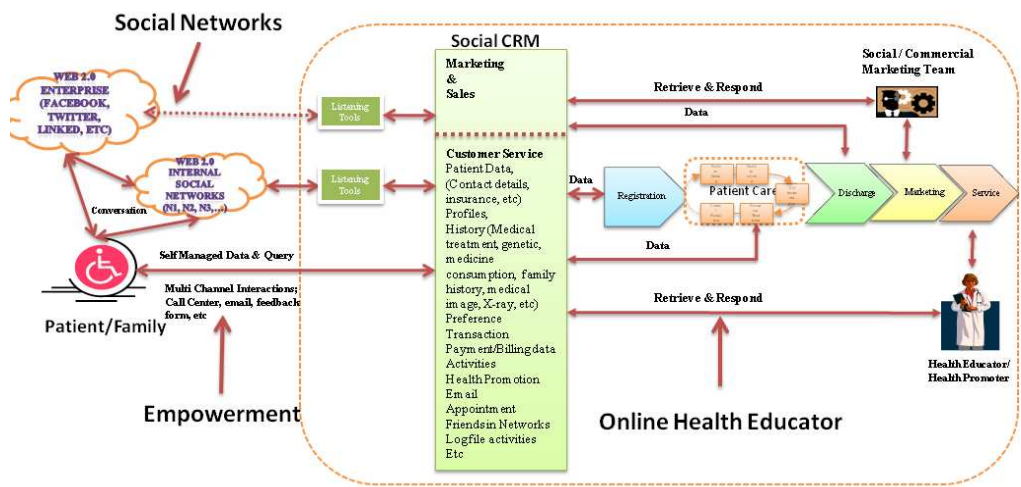

**Fig 3. Proposed Model of Social CRM in Healthcare**

The framework is developed from Enterprise Social Networks, Internal Social Networks, Listening tool interfaces, Social CRM systems within healthcare provider, and healthcare value configuration (value chain and value shop).

Social Networks refers to the Web 2.0 technology that patient or his/her families may join any of them. It differentiates two social networks linkages to the patient or his/her family; they are Enterprises Social Networks and Internal Social Networks. The Enterprises Social Networks refers to external and popular Web 2.0 applications such as Facebook, Twitter, LinkedIn, MySpace, Friendster, etc which patient may belong to any of those for interaction. The dashed line connected enterprises social networks and CRM systems mean that none of those networks have control over the others directly, but constructive conversation and information from enterprises social networks should be captured for creating strategy, innovation, better service and at the same time responds accurately.

Lesson learnt from case 1, the hospital was not proficient in capturing the message from customers at social networks because they did not consider it in their relationship management strategy that the customers have changed and they made conversation, judging hospital's value, criticizing their services at those networks which led to distrust towards the hospital's allegation and jeopardized the business for the long run (Anshari and Almunawar, 2011).

Further, it proposes Internal Social Networks that operated, managed, and maintained within healthcare's infrastructure. This is more targeted to internal patients/families within the healthcare to have conversation between patients/family within the same interest or health problem/ illness. For example, a diabetic patient is willing to share his/her experiences, learning, and knowledge with other diabetic patients. Since patient/family who generates the contents of the Web, it can promote useful learning center for others, not only promoting health among each others, but also it could be the best place supporting group and sharing their experiences related to all issues such as; how the healthcare does a treatment, how much it will cost them, what insurance accepted by healthcare, how is the food and nutrition provided, etc. Therefore, this is generic group that will grow depends on the need of patients in that healthcare. For instant, N1 is internal social networks for Diabetic, N2 is for Cancer, N3 is for hearth disease, and so on. Creating Internal Social Networks is part of the strategy to isolating problem into small space or more focus to the local's problem so it can be easily monitored and solved. Moreover, this strategy will promote loyalty of customers to keep using service from the healthcare.

In general, the aim to put together linkage of internal and external social networks are to engage patients and export ideas, foster innovations of new services, quick response/feedback for existing service, and technologies from people inside and outside organization. Both provide a range of roles for patient or his/her family. The relationships can create emotional support, substantial aid and service, influence, advice, and information that a person can use to deal with a problem. In addition, listening tool between Social Networks and CRM systems (see figure 6.) is mechanism to capture actual data from social media and propagates this information forward to the CRM. This tool should be capable to filter noise (level of necessity for business process) from actual data that needs to be communicated to CRM.

Since this strategy was absent in the Prita's case, the hospital was late to realize that the patient dissatisfied with the service from beginning, and the hospital assumed that everything was fine, until she communicated her dissatisfaction through her social networks. Responding this problem, the hospital should isolate the internal problem like dissatisfaction of patient by quick response to resolve the issue before it gets bigger and uncontrolled. The Internal Social Networks could be solution to prevent the same problem in the future.

Social CRM empowers patient/family to have the ability in controlling his own data. Once patient/family registers to have service from healthcare provider, it will enable them to have personalized e-health systems with Social CRM as the frontline of the system. The system will authorize for each patient then; the authorization and self-managed account/service are granted to access all applications and data offered by the systems. This authorization is expected to be in the long run since the information and contents continue to grow. Technical assistant is available through manual or health informatics officer (just like any other customer service in business/organization) who stand by online assisting patient/family in utilizing the systems especially for the first timer. Furthermore, since all the information (medical records) can be accessed online everywhere and anytime, it will enable collaborative treatment from telemedicine. Empowerment is the alternative solution for the case 2, there is a demand from the customer to be able to manage any activities that they believe can improve efficiency and effectiveness of the healthcare process. And the empowerment can be strategy for the healthcare organization in reducing complains from the customers.

Consider this scenario; while we go to physician for diagnose, sometimes there is tradeoff between time allocated each patient and comprehensiveness of the diagnosing process. Long queue patient waits for consultation make healthcare provider to be able to allocate time wisely for each patient. Within the constraint of consultation time yet physician is able to conduct diagnosis efficiently and effectively. The system supports the customer service because it helps both healthcare provider and patient in diagnose activity. The physician will have complete information, knowledge, and saving a lot of time to learn about patient history because patient participate in the detailing his medical records data through the system, and patient benefits from quality of diagnoses' time because his medical records are overviewed in full scene. In other words, it can provide better customer service to meet patient's expectation and improve quality of consultation time. The physician is expected to have comprehensive view of the patient's history before diagnosing or analyzing consulted symptoms. This can be achieved because physician will be able to observe the report of patient's medical history such as last medicine consumption, previous diagnoses, lab result, activities suggested by health educator, etc. In addition, by empowering patients with medical data and personalized e-health, the healthcare needs to provide officer in duty (health educator/ health promoter) in order to interpret medical data or respond online query/consultation. The officer in duty is required to have an ability to interpret medical data and also familiar with the technical details of the systems.

Social CRM functionalities like previous version are composed from marketing, sales, and customer service which operated to achieve business strategy of healthcare organization. For example, marketing's strategy should accommodate social marketing to promote public health and commercial marketing to acquire more customers coming for services. Customer service will offer distinct value for each activity. The different from the traditional CRM, the state to empowering for self managed data and authorization will encourage patient willingly to provide full data without hesitation. More data provided more information available for the sake of analyzing for the interest of marketing, sales, and customer service.

The own unique characteristics value creation by adopting Social CRM is ability to generate contents from both parties either from healthcare and also patients. Social CRM in healthcare is providing value-added services to patients like openness of medical records, improving patient loyalty, creating better healthcare-patient communication, improving brand image and recognition, and self managed data which will improve health literacy to reduce economic burden for society to the whole.

The framework above proposes and combines the concept of value chain and value shop. As discussed in previous section, the raw data arrives in one state, and leave in another state. The patient enters ill and leaves well (hopefully). The activities of value chain are; arriving from registration, patient care, discharge, marketing, and service—producing data at respective state. However, in the process of patient care is elaborated according to the value shop where value is created by mobilizing resources and activities to resolve a particular patient's problem. The five generic categories of primary value shop activities; Problem-finding and acquisition, choosing the overall approach to solving the problem execution, and control & evaluation

Furthermore, the framework accommodates Social/Commercial Marketing team. Social marketing is more prevalence to the government healthcare that operates as an agent of the public a7 t large. Campaign of healthy life for the healthcare is example of Social Marketing. It is not intended for commercial benefit for short term but it is beneficial for the community. On the other hand, commercial marketing is standard marketing strategy exist for any business entities. Both are acting in responds to the public demands like social networks, Mailing list, blog, etc.

The other feature of the model is robustness of systems because more applications/services will be added as characteristics of Web 2.0. Some of the features that available to the user are; updating personal data, Medical Records & History (medical treatment received, medicine consumption history, family illness history, genetic, medical imaging, x-ray, etc), Preference services, Transaction, Payment/Billing data, Activities, Personal Health Promotion and Education, Email, Appointment, Friend in networks, forums, chatting, etc.

In summary, table 2 below shows comparison between CRM 1.0 and CRM 2.0 in the healthcare environment. Social CRM presents openness which all activities involve with patient recorded on systems and patients/families are able to access them online. The privileges to access medical records, patient personal data, appointment with physician, scheduling, and any other features of Web 2.0 added to establish conversation, convenience, and creating trust to the service is innermost for healthcare that employ Social CRM.

**Table 2: Comparison CRM 1.0 and CRM 2.0 in Healthcare Organization**

| Type | CRM 1.0 | CRM 2.0 |
|---|---|---|
| **Relationship** | Healthcare provider and patients | Healthcare provider, patients, public, & social networks |
| **Connection** | Limited view of patient's preference | Better understanding of patient's preference |
| **Generated value** | Targeted message generates value | Conversation generates value to provider & patients |
| **Accessibility** | Limited access to medical records | Patient has better control to his own data |
| **Technology** | Mostly based on Web 1.0 technology & provider generates contents. | Based on Web 2.0, provider & patients are able to generate contents |

From patient perspective, their health literacy is expected to improve by the time they have better knowledge of their own health status, and for the healthcare organization is expected to be long lasting relationship since they always in need to access the service. It is justified that the empowerment, online health educator, and discussion that take place in social network in Social CRM is critical to the success of e-health.

The adoption of Social CRM in e-health systems helps to prevent any dispute and avoiding conflict between healthcare organization and patient. The hospital needs to understand that behavior and expectation of patients continue to change eventually. We propose that Social CRM framework as alternative solution to the hospital. Late responds may affects survival ability of the hospital in the long run and jeopardize the business due to loosing of the trust towards the service. Therefore, the hospital should perform re-engineering of processes to adapt their CRM strategy and tool in order to acquire potential customer coming for the service. The framework will give prospect to acquire, retain, and extend relationship with their patients.

## Conclusion

Managing relationship with patients is one of the most remarkable aspects in healthcare service due to the complexity of the process. The fact that patient expectations in healthcare are high, create a serious challenge for healthcare providers to meet those expectation. A new paradigm has appeared in CRM systems namely Social CRM or CRM 2.0 as a result of the development of Web 2.0 technology. By inheriting features from the Web 2.0 technology, Social CRM offers new outlook either from patient or healthcare. Some of the features offered by Social CRM framework are robustness of the systems, ingenuousness/openness of information sharing, and closeness of relationship between patient-healthcare and patient with others. The systems create value in each activity to the customer. And those values will make the healthcare a better service than before. Moreover, it empowers patients with the data accessibility in returns of loyalty and trust relationship with their patients.

The use of Social CRM in e-health system is equivalent to bringing patient expectation aligned with fashion of ICT in actual healthcare services. It offers new outlook either from patient or healthcare organization, and how they structure inter-relation between three distinct domains of objects; customer's expectation, advancement of ICT, and healthcare services. Each domain has unique features and characteristics which failing to respond appropriately may affect to business survivability and customer dissatisfaction. These features are empowerment, online social networks, and online health education. Though the framework is mainly proposed in the healthcare's environment, but it is generic model which bring promise to service provider by considering Social CRM in creating business strategy, supported by a Web as platform, business processes, and social characteristics to engage the customer in a collaborative conversation in order to provide mutually beneficial value in a trusted and transparent business environment